\documentclass[twocolumn,preprintnumbers,amsmath,amssymb,secnumarabic,amssymb, nobibnotes, aps, pra]{revtex4-1}
\usepackage{times}
\usepackage{graphicx}
\usepackage{dcolumn}
\usepackage{bm}
\usepackage{braket}
\usepackage{mathrsfs}
\begin{document}
\title{Effective unidirectional pumping for steady-state amplification without inversion.}
\author{Pankaj K. Jha}\email{Email: pkjha@physics.tamu.edu}
\affiliation{Department of Physics and Astronomy, Texas A\&M University, College Station, TX-77843 USA}
\begin{abstract}
\noindent We discuss an opportunity to achieve amplification without inversion in three-level cascade scheme using an effective unidirectional pumping via bidirectional incoherent pump. Analytical solution to the population and the coherence are obtained in the steady-state regime. With a proper choice of the parameters, obtained here, the possibility for amplification without inversion is presented.
\noindent
\end{abstract}
\date{\today}
\maketitle
\section{Introduction}
It is well known that a conventional laser requires population inversion on the lasing transition. But in late 80's it was proposed ~\cite{LWI1,LWI2,LWI3} and later demonstrated experimentally~\cite{LWI4,LWI5,LWI12} that this condition of population inversion does not hold true in general when more than two levels are involved in the interaction.  In a typical three-level configuration, it is possible to break the detail balance and cancel or suppress absorption while keeping the stimulated emission intact. This is the basis of lasing without inversion (LWI). Many schemes for lasing without inversion has been proposed~\cite{LWI1,LWI2,LWI3,LWI6,LWI7,LWI8,LWI9,LWI10,LWI11,LWI13}\,(see also the overviews\cite{R1,R2,R3,R4}). The key to all the schemes proposed and realized is quantum coherence and interference effects. Indeed the generation of short-wavelength radiation is one of the core application of lasing without inversion where fast spontaneous relaxation times makes the realization of population inversion difficult. Necessary condition for amplification without inversion (AWI) has been obtained and AWI has been related to inversion in field reservoir\cite{Y1}, AWI has been shown for a medium in a thermal equilibrium \cite{Y2} (without any inversion including inversion in field reservoir), LWI without external coherent drive~\cite{Y21}, LWI due to coherence excited via spontaneous decay \cite{Y3}, mechanisms of LWI have been discussed in \cite{Y4,Y5}, extension to free electron lasers \cite{Y6}, to gamma-ray region \cite{Y7}. The quest for X-ray lasers, have motivated researchers in this area, to explore different schemes of lasing with or without inversion~\cite{XR}.

In this paper we start with a brief review of the steady-state gain without population inversion in a three-level medium (cascade configuration) with incoherent pumping (symmetric and bidirectional) between the ground state $|a\rangle$ and the excited level $|c\rangle$ (as shown in Fig. 1(a)). Similar scheme with asymmetric, symmetric and unidirectional incoherent pumping has been addressed extensively~\cite{Shuker1,Shuker2} in context with lasing on the X-Ray transition of Ar$^{8+}$ at wavelength around $26nm$. When we have a symmetric incoherent pump, though the lasing transition never reaches population inversion, it also never exhibits amplification and shows absorption. This undesirable outcome can be overcome if we introduce a new level $|p\rangle$ (as shown in Fig. 1(b)). We show that if the symmetric and bi-directional pump is introduced in the transition $|p\rangle \leftrightarrow |b\rangle$, the probe transition $|b\rangle \leftrightarrow |p\rangle$ can exhibit gain without population inversion. The main result of this paper is the Fig. 2(c) and the relation Eq.(\ref{EQ19}).

The paper is organized as follows. In section II, we briefly review the gain(absorption) profile of a three-level medium in cascade configuration in steady-state regime with incoherent coupling (symmetric and bidirectional) between the ground state $|a\rangle$ and the excited level $|c\rangle$. In section III, we show that by introducing a fourth level $|p\rangle$, coupled to the ground state by a bidirectional incoherent pump, the probe transition $|a\rangle \leftrightarrow |b\rangle$ can exhibit amplification without population inversion [see Fig 2(c)] for a proper choice of the parameters. We also show some light on the temporal behavior and the effect of probe detuning $(\Delta_{b})$ on the gain for the four-level model. In Appendix A, we have briefly discussed the three-level model in $\Lambda-$configuration and show that the system can exhibit gain even in the presence of a symmetric and bi-directional pump between the lower two levels. 
\section{Three-Level Model}
\noindent We consider a three-level model as shown in Fig. 1(a). The transition $|c\rangle \leftrightarrow |a\rangle $ is driven by a coherent driving field and the transition $|a\rangle \leftrightarrow |b\rangle $ is excited by a weak probe field. The population in transition $|c\rangle \leftrightarrow |a\rangle $ is also exchanged using an incoherent symmetric and bi-directional pump at a rate $\lambda$.  The atom-field Hamiltonian in the interaction picture with rotating-wave approximation can be written as~\cite{MOS} $(\hbar=1)$
\begin{equation}\label{EQ1}
\mathcal{V}=\Delta_{a}|c\rangle\langle c|-\Delta_{b}|b\rangle\langle b|-\left(\Omega_{b}\left |a \rangle \langle b \right |+\Omega_{a}\left |c \rangle \langle a \right | +\text{H.c}\right).
\end{equation}
\begin{figure}[t]
\centerline{\includegraphics[height=4.4cm,width=0.47\textwidth,angle=0]{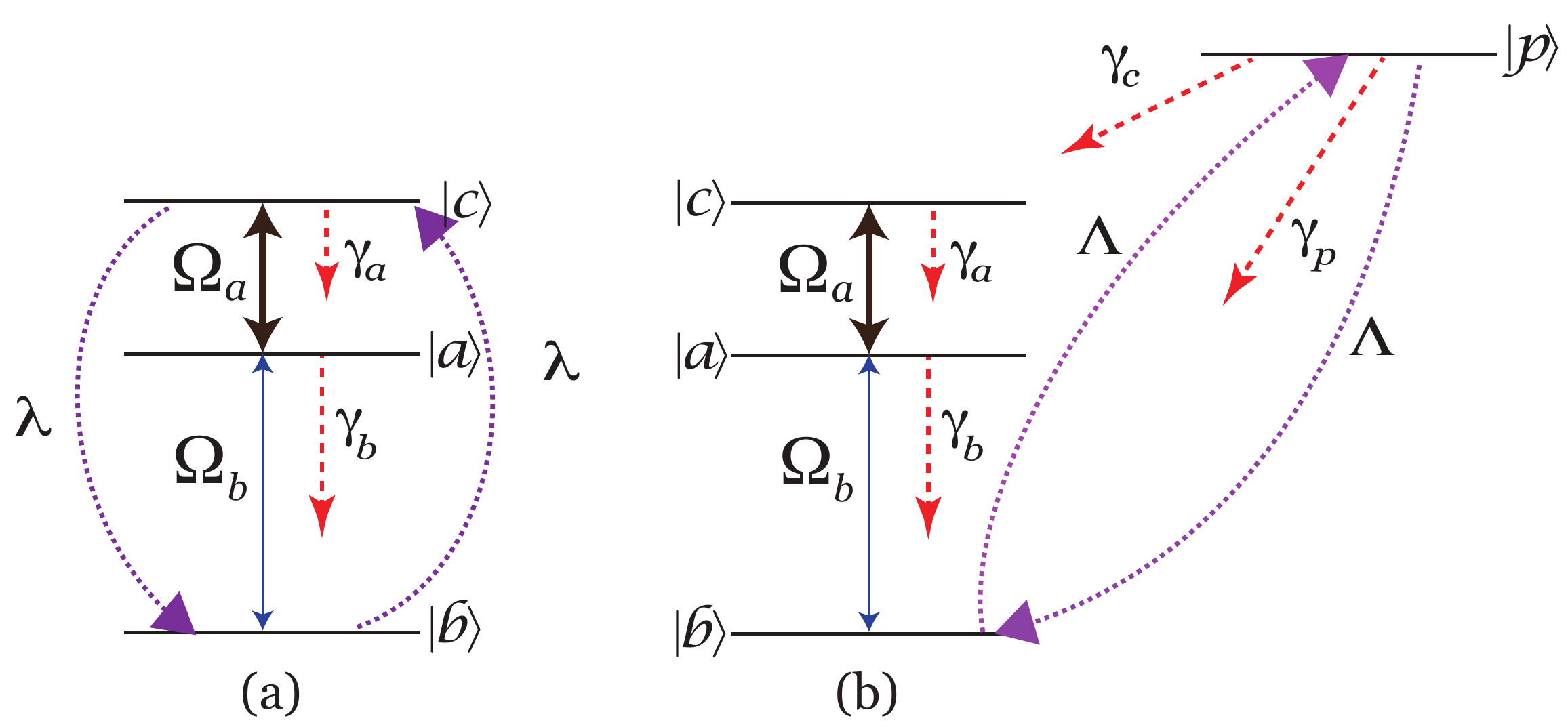}}
 \caption{Level diagram for the (a) three-level and (b) four-level model. The spontaneous decay rates $|a\rangle \rightarrow |b\rangle$, $|c\rangle \rightarrow |a\rangle$, $|p\rangle \rightarrow |c\rangle$ and $|p\rangle \rightarrow |b\rangle$ are give by $\gamma_{b}, \gamma_{a}, \gamma_{c},\gamma_{p}$ respectively. We have introduced an incoherent bi-directional pump in the transition $|b\rangle \leftrightarrow |c\rangle$ for the three-level model and in the transition $|b\rangle \leftrightarrow |p\rangle$ for the four-level medium. The pumping rates are $\lambda$ and $\Lambda$ for the three and four-level model respectively. $\Omega_{a}$ and $\Omega_{b}$ are the drive and probe field Rabi frequencies.}
\end{figure}
\noindent Here $\Omega_{a}$ and $\Omega_{b}$ are the Rabi frequencies of the driving and the probe field respectively. We define the detuning $\Delta_{a} =\omega_{ca}-\nu_{a}$ and $\Delta_{b}=\omega_{ab}-\nu_{b}$~\cite{D}. In this model the spontaneous decay in the channels $ab$, $ca$ are quantified by the parameter $\gamma_{b}$,  $\gamma_{a}$ respectively. Incorporating the decay rates, the equation of motion for the atomic density matrix is given as  $(\hbar=1)$
\begin{equation}\label{EQ2}
\begin{split}
&\frac{\partial \varrho}{\partial t}=-i[\mathcal{V},\varrho]+\frac{\gamma_{b}}{2}\left([\sigma_{b},\varrho\sigma_{b}^{\dagger}]+[\sigma_{b}\varrho,\sigma_{b}^{\dagger}]\right)\\
&+\frac{\gamma_{a}}{2}\left([\sigma_{a},\varrho\sigma_{a}^{\dagger}]+[\sigma_{a}\varrho,\sigma_{a}^{\dagger}]\right)+\frac{\lambda}{2}\left([\sigma_{\lambda},\varrho\sigma_{\lambda}^{\dagger}]+[\sigma_{\lambda}\varrho,\sigma_{\lambda}^{\dagger}]\right)\\
&+\frac{\lambda}{2}\left([\sigma^{\dagger}_{\lambda},\varrho\sigma_{\lambda}]+[\sigma_{\lambda}^{\dagger}\varrho,\sigma_{\lambda}]\right),
 \end{split}
\end{equation}
where the atomic lowering ($\sigma_{i}$) and rising operators ($\sigma^{\dagger}_{i}$) are defined as
\begin{equation}\label{EQ3}
\begin{split}
 \sigma_{a}=\left |a \rangle \langle c \right |,\, \sigma_{b}=\left |b \rangle \langle a \right |,\,  \sigma_{\lambda}=\left |b \rangle \langle c \right |, \\
 \sigma_{a}^{\dagger}=\left |c \rangle \langle a \right |,\, \sigma_{b}^{\dagger}=\left |a \rangle \langle b \right |,\, \sigma_{\lambda}^{\dagger}=\left |c \rangle \langle b \right |.
 \end{split}
\end{equation}
\noindent The equations of motion for the density matrix element $\varrho_{ab}, \varrho_{cb}$ and $\varrho_{ca}$ is given by (for real $\Omega_{a}, \Omega_{b}$)
\begin{subequations}\label{EQ4}
\begin{align}
\frac{\partial \varrho_{ab}}{\partial t}&=-\Gamma_{ab}\varrho_{ab}-i\Omega_{b}(\varrho_{aa}-\varrho_{bb})+i\Omega_{a}\varrho_{cb},\\
\frac{\partial \varrho_{cb}}{\partial t}&=-\Gamma_{cb}\varrho_{cb}+i\Omega_{a}\varrho_{ab}-i\Omega_{b}\varrho_{ca},\\
\frac{\partial \varrho_{ca}}{\partial t}&=-\Gamma_{ca}\varrho_{ca}-i\Omega_{a}(\varrho_{cc}-\varrho_{aa})-i\Omega_{b}\varrho_{cb}.
\end{align}
\end{subequations}
Here $\Gamma_{ab}=(\gamma_{b}+\lambda)/2+i\Delta_{b}, \Gamma_{ca}=(\gamma_{a}+\gamma_{b}+\lambda)/2+i\Delta_{a}$ and $\Gamma_{cb}=\gamma_{a}/2+\lambda+i\left(\Delta_{a}-\Delta_{b}\right)$. For a weak probe field, a first-order solution for $\varrho_{ab}$ (which determines the gain/absorption of the probe field) can be found in the steady state
\begin{equation}\label{EQ5}
\varrho^{(1)}_{ab}=-i\Omega_{b}\left\{\frac{\left[\varrho^{(0)}_{aa}- \varrho^{(0)}_{bb}\right]\Gamma_{cb}\Gamma_{ca}+\left[\varrho^{(0)}_{cc}- \varrho^{(0)}_{aa}\right]\Omega_{a}^{2}}{\left(\Gamma_{cb}\Gamma_{ab}+\Omega_{a}^2\right)\Gamma_{ca}}\right\},
\end{equation}
where $\varrho^{(0)}_{ll}$ is the zeroth-order population in the level $|l\rangle$. For resonant interaction$\left(\Delta_{a,b}=0\right)$, if we look at Eq.(5), the imaginary part of $\varrho_{ab}$ denoted by  $\Im[\varrho_{ab}]$ has two contributing terms. While the first term is proportional to the population inversion $\left(\propto \left[\varrho^{(0)}_{aa}- \varrho^{(0)}_{bb}\right]\right)$ in the probe transition, the second term is proportional to population inversion $\left(\propto \left[\varrho^{(0)}_{cc}- \varrho^{(0)}_{aa}\right]\right)$ in the driving transition. In case of two-level model $\Omega_{a}=0$, we need $\varrho^{(0)}_{aa}> \varrho^{(0)}_{bb}$ i.e population inversion in probe transition for amplification~\cite{Mollow}. For three-level model it is the second term which provides the necessary conditions required for lasing without population inversion.
\subsection{Steady-state analysis}
\noindent The equation of motion of the density matrix elements $\varrho_{ll}$ are given by
\begin{subequations}\label{EQ6}
\begin{align}
\frac{\partial \varrho_{aa}}{\partial t}&=-\gamma_{b}\varrho_{aa}+\gamma_{a}\varrho_{cc}-2\Omega_{a}\Im[\varrho_{ca}]+2\Omega_{b}\Im[\varrho_{ab}],\\
\frac{\partial \varrho_{bb}}{\partial t}&=\gamma_{b}\varrho_{aa}-\lambda \rho_{bb}+\lambda\varrho_{cc}-2\Omega_{b}\Im[\varrho_{ab}],\\
\frac{\partial \varrho_{cc}}{\partial t}&=-(\gamma_{a}+\lambda)\varrho_{cc}+\lambda\varrho_{bb}+2\Omega_{a}\Im[\varrho_{ca}].
\end{align}
\end{subequations}
The exact steady-state solution for the coupled Eqs.(\ref{EQ6}) is complex even for resonant interaction $\left(\Delta_{a,b}=0\right)$\cite{J1,J2,J3}. To obtain solution in compact analytical form, we employed the zeroth order approximation in the probe field and obtained for the steady-state populations
\begin{subequations}\label{EQ7}
\begin{align}
\varrho^{(0)}_{aa}&=\frac{\lambda(\gamma_{a}\Gamma_{ca}+2\Omega^{2}_{a})}{\mathcal{M}},\\
\varrho^{(0)}_{bb}&=\frac{\lambda(\gamma_{b}\Gamma_{ca}+2\Omega^{2}_{a})+\gamma_{b}(\gamma_{a}\Gamma_{ca}+2\Omega^{2}_{a})}{\mathcal{M}},\\
\varrho^{(0)}_{cc}&=\frac{\lambda(\gamma_{b}\Gamma_{ca}+2\Omega^{2}_{a})}{\mathcal{M}},
\end{align}
\end{subequations}
where $\mathcal{M}=\lambda(2\gamma_{b}\Gamma_{ca}+4\Omega^{2}_{a})+(\gamma_{b}+\lambda)(\gamma_{a}\Gamma_{ca}+2\Omega^{2}_{a}).$ Using Eqs.(\ref{EQ5},\ref{EQ7}), we obtain the first-order solution for $\varrho^{(1)}_{ab}$
\begin{equation}\label{EQ8}
\begin{split}
\varrho^{(1)}_{ab}=-i\Omega_{b}\left\{\frac{-\Gamma_{cb}\Gamma_{ca}\left[\gamma_{a}(\gamma_{b}-\lambda)+\lambda\gamma_{b}\right]}{\left(\Gamma_{cb}\Gamma_{ab}+\Omega^{2}_{a}\right)\mathcal{M}}\right.\\
\left. +\frac{\left[\gamma_{b}(\lambda-2\Gamma_{cb})-\gamma_{a}\lambda\right]\Omega^{2}_{a}}{\left(\Gamma_{cb}\Gamma_{ab}+\Omega^{2}_{a}\right)\mathcal{M}}\right\}.
\end{split}
\end{equation}
\begin{figure*}[htb]
\includegraphics[height=5.6cm,width=0.33\textwidth,angle=0]{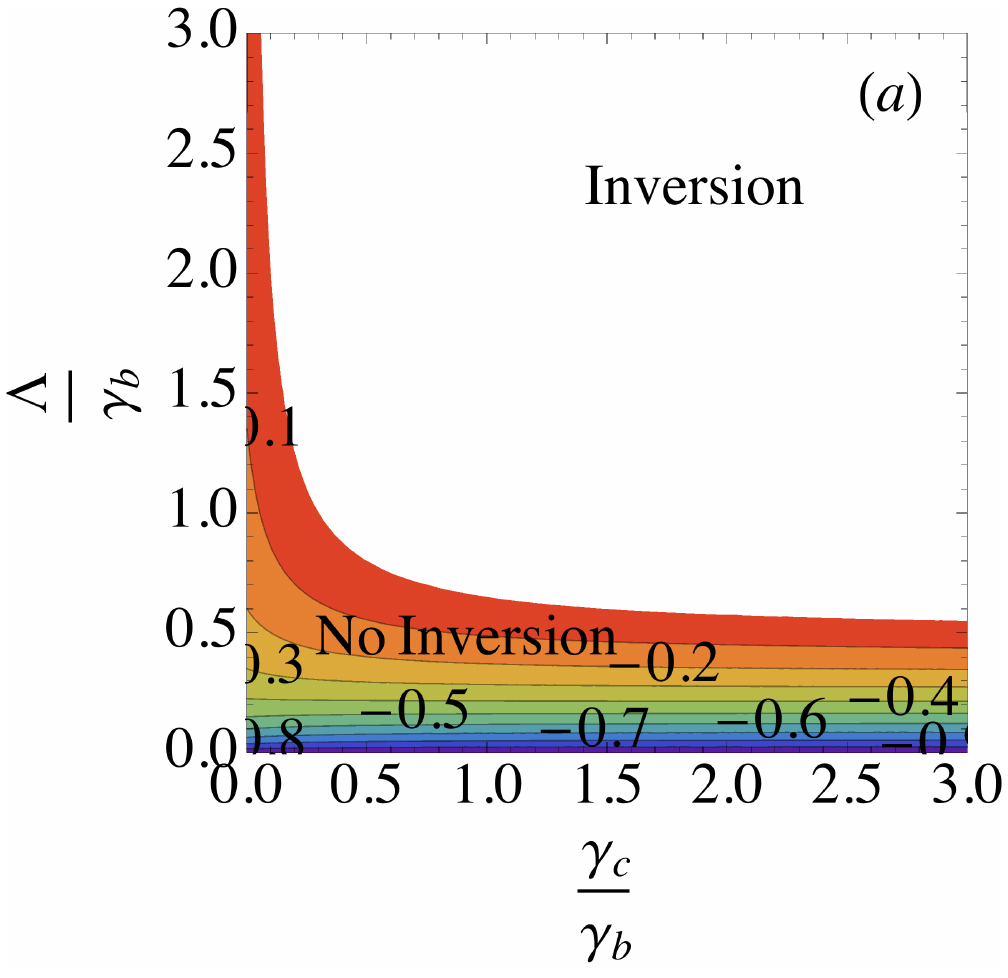}
\includegraphics[height=5.6cm,width=0.32\textwidth,angle=0]{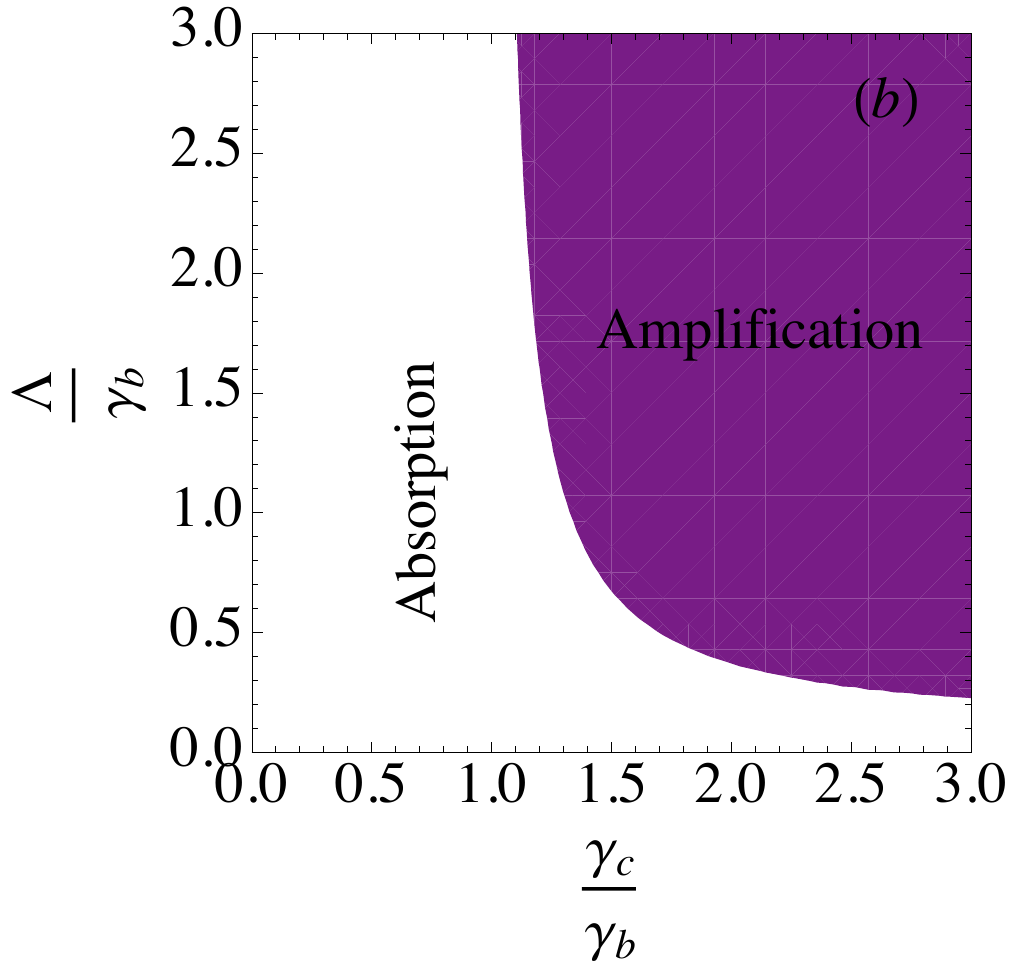}
\includegraphics[height=5.6cm,width=0.33\textwidth,angle=0]{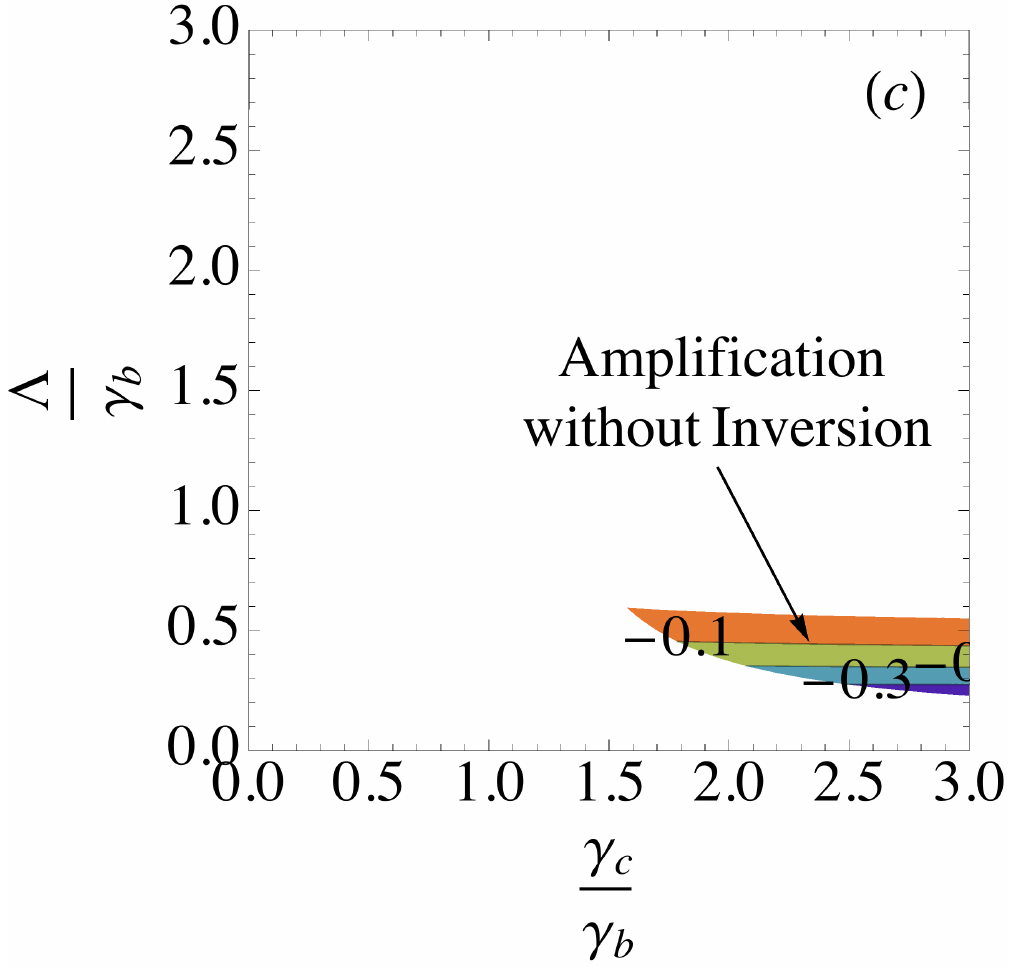}
 \caption{ContourPlots for the steady-state population inversion i.e $\left(\varrho^{(0)}_{aa}+\varrho^{(0)}_{cc}+\varrho^{(0)}_{pp}-\varrho^{(0)}_{bb}\right)$ and coherence in the probe transition $|a\rangle \leftrightarrow |b\rangle$ with a symmetric and bi-directional incoherent pump in the transition $|b\rangle \leftrightarrow |p\rangle$ for the four-level model Fig 1(b). (a) Plot of the population inversion against the decay rate $\gamma_{c}/\gamma_{b}$ and pumping rate $\Lambda/\gamma_{b}$. The shaded region corresponds to no-inversion while the unshaded region shows the values of the parameters $\gamma_{c}/\gamma_{b}$ and $\Lambda/\gamma_{b}$ for which we have an inverted system in steady-state. (b) Similar plot of the $\Im[\varrho_{ab}]$. Here the shaded region corresponds to gain ($\Im[\varrho_{ab}]<0$) while the unshaded region will have absorption in the probe transition ($\Im[\varrho_{ab}]>0$). (c) Plot of the region i.e range of the parameters $\gamma_{c}/\gamma_{b}$ and $\Lambda/\gamma_{b}$  for which amplification without population inversion in the probe transition can be achieved. For numerical simulation we took $\Omega_{a}=10, \Omega_{b}=0.01, \gamma_{a}=0.01, \gamma_{p}=0.3, \gamma_{b}=1$.}
\end{figure*}
The condition for non-inversion $\varrho^{(0)}_{aa}+\varrho^{(0)}_{cc}<\varrho^{(0)}_{bb}$\,~\cite{Y1,SH1} gives
\begin{equation}\label{EQ9}
\frac{(\lambda-\gamma_{b})(\gamma_{a}\Gamma_{ca}+2\Omega^{2}_{a})}{\mathcal{M}} <0.
\end{equation}
\noindent From Eq.(\ref{EQ9}), we see that for a symmetric and bidirectional incoherent pumping in the transition $|c\rangle \leftrightarrow |a\rangle$, population is never inverted in steady state if $\lambda <\gamma_{b}$ i.e rate of incoherent pump should be less than the rate of radiative decay from $|a\rangle \rightarrow |b\rangle$. We know that the imaginary part of $\varrho_{ab}$ i.e $\Im[\varrho_{ab}]$ governs the gain(absorption) in the probe transition. For gain(absorption) in the $|a\rangle \leftrightarrow |b\rangle$ transition we need $\Im[\varrho_{ab}] <0(>0)$. From Eq. (\ref{EQ8}) we obtain, in the limit $\gamma_{b} \gg \lambda \gg \gamma_{a}$
\begin{equation}\label{EQ10}
\Im[\varrho_{ab}] \propto \gamma_{b}\lambda\left[ \gamma_{b}\lambda/2+\Omega^{2}_{a}\right]\Omega_{b}\, >0.
\end{equation}
As $\Im[\varrho_{ab}] >0$, the probe transition will always exhibit absorption in steady-state. To conclude, in the presence of a bi-directional symmetric incoherent pump between $|c\rangle \leftrightarrow |a\rangle$ levels, the probe transition will never exhibit gain in $\Xi$-configuration. However this can be overcome in case of an asymmetric bi-directional pump~\cite{Shuker1,Shuker2}. In the next part of this paper we studied a four-level model with bi-directional incoherent pump between the ground state $|b\rangle $ and an additional fourth level $|p\rangle$ as shown in Fig. 1(b) and discuss the opportunity to observe amplification without inversion.
\section{Four-Level Model}
We will now consider a four-level model shown in Fig. 1(b), the atom-field interaction in the interaction picture with rotating-wave approximation is also given by Eq. (\ref{EQ1}).  The spontaneous decay in the channel $pc$ and $pb$ are given by the parameters $\gamma_{c}$ and $\gamma_{p}$ respectively. Incorporating these additional decay rates, the equation of motion for the atomics density matrix is
\begin{equation}\label{EQ11}
\begin{split}
\frac{\partial \varrho}{\partial t}&=-\frac{i}{\hbar}[\mathcal{V},\varrho]+\frac{\gamma_{b}}{2}\left([\sigma_{b},\varrho\sigma_{b}^{\dagger}]+[\sigma_{b}\varrho,\sigma_{b}^{\dagger}]\right)\\
&+\frac{\gamma_{a}}{2}\left([\sigma_{a},\varrho\sigma_{a}^{\dagger}]+[\sigma_{a}\varrho,\sigma_{a}^{\dagger}]\right)+\frac{\gamma_{c}}{2}\left([\sigma_{c},\varrho\sigma_{c}^{\dagger}]+[\sigma_{c}\varrho,\sigma_{c}^{\dagger}]\right)\\
&+\frac{\tilde{\gamma}_{p}}{2}\left([\sigma_{p},\varrho\sigma_{p}^{\dagger}]+[\sigma_{p}\varrho,\sigma_{p}^{\dagger}]\right)+\frac{\Lambda}{2}\left([\sigma^{\dagger}_{p},\varrho\sigma_{p}]+[\sigma_{p}^{\dagger}\varrho,\sigma_{p}]\right),
 \end{split}
\end{equation}
where $\tilde{\gamma}_{p}=\gamma_{p}+\Lambda$ and the atomic lowering ($\sigma_{i}$) and rising operators ($\sigma^{\dagger}_{i}$) are defined as
\begin{equation}\label{EQ12}
\begin{split}
 \sigma_{b}=\left |b \rangle \langle a \right |, \sigma_{b}^{\dagger}=\left |a \rangle \langle b \right |\,;\,  \sigma_{a}=\left |a \rangle \langle c \right |, \sigma_{a}^{\dagger}=\left |c \rangle \langle a \right |, \\
  \sigma_{c}=\left |c \rangle \langle p \right |, \sigma_{c}^{\dagger}=\left |p \rangle \langle c \right |\,;\,  \sigma_{p}=\left |b \rangle \langle p \right |, \sigma_{p}^{\dagger}=\left |p \rangle \langle b \right |.
 \end{split}
\end{equation}
The equation of motion for the density matrix elements $\varrho_{ab},\varrho_{cb}$ and $\varrho_{cb}$ takes the form given by Eq.(\ref{EQ4}) with the parameters $\Gamma_{ab}=(\gamma_{b}+\Lambda)/2 +i\Delta_{b},\, \Gamma_{ca}=(\gamma_{a}+\gamma_{b})/2 +i\Delta_{a}$ and $\Gamma_{cb}=(\gamma_{a}+\Lambda)/2+i(\Delta_{a}-\Delta_{b})$. The first-order solution for $\varrho_{ab}$ is given by Eq.(\ref{EQ5}). Eqs. (\ref{EQ4},\ref{EQ5}) are quite general equations, for the Hamiltonian governed by Eq.(\ref{EQ1}).
\subsection{Steady-state analysis}
The equation of motion of the density matrix elements $\varrho_{ll}$ are given by
\begin{subequations}\label{EQ13}
\begin{align}
\frac{\partial \varrho_{aa}}{\partial t}&=-\gamma_{b}\varrho_{aa}+\gamma_{a}\varrho_{cc}-2\Omega_{a}\Im[\varrho_{ca}]+2\Omega_{b}\Im[\varrho_{ab}],\\
\frac{\partial \varrho_{bb}}{\partial t}&=\gamma_{b}\varrho_{aa}-\Lambda \rho_{bb}+\tilde{\gamma}_{p}\varrho_{pp}-2\Omega_{b}\Im[\varrho_{ab}],\\
\frac{\partial \varrho_{cc}}{\partial t}&=-\gamma_{a}\varrho_{cc}+\gamma_{c}\varrho_{pp}+2\Omega_{a}\Im[\varrho_{ca}],\\
\frac{\partial \varrho_{pp}}{\partial t}&=\Lambda \varrho_{bb}-(\gamma_{c}+\tilde{\gamma}_{p})\varrho_{pp}.
\end{align}
\end{subequations}
In steady-state we obtain the zeroth-order population [see Appendix for calculations]
\begin{subequations}\label{EQ14}
\begin{align}
\varrho^{(0)}_{aa}&=\frac{\gamma_{c}\Lambda(\gamma_{a}\Gamma_{ca}+2\Omega^{2}_{a})}{\mathcal{D}},\\
\varrho^{(0)}_{bb}&=\frac{\gamma_{b}(\gamma_{c}+\tilde{\gamma}_{p})(\gamma_{a}\Gamma_{ca}+2\Omega^{2}_{a})}{\mathcal{D}},\\
\varrho^{(0)}_{cc}&=\frac{\gamma_{c}\Lambda(\gamma_{b}\Gamma_{ca}+2\Omega^{2}_{a})}{\mathcal{D}},\\
\varrho^{(0)}_{pp}&=\frac{\gamma_{b}\Lambda(\gamma_{a}\Gamma_{ca}+2\Omega^{2}_{a})}{\mathcal{D}},
\end{align}
\end{subequations}
where, $\mathcal{D}=2\Omega^{2}_{a}\left[2\Lambda\gamma_{c}+\gamma_{b}(\Lambda+\gamma_{c}+\tilde{\gamma}_{p})\right] +\gamma_{c}\gamma_{b}\Lambda\Gamma_{ca}+\gamma_{a}\Gamma_{ca}\left[\gamma_{c}\Lambda+\gamma_{b}(\Lambda+\gamma_{c}+\tilde{\gamma}_{p})\right]$.
\subsection{Gain condition}
From the condition for non-inversion $\varrho^{(0)}_{aa}+\varrho^{(0)}_{cc}+\varrho^{(0)}_{pp}<\varrho^{(0)}_{bb}$ we obtain (for large $\Omega_{a} \gg \gamma_{a}, \gamma_{b},\gamma_{c}$)
\begin{equation}\label{EQ15}
\gamma_{c}(2\Lambda-\gamma_{b}) <\gamma_{b}\gamma_{p}.
\end{equation}
\noindent We can easily see from Eq.(\ref{EQ15}), for $\Lambda <\gamma_{b}/2$, the system will never shows population inversion in steady state [see Fig.2 (a)]. Using Eq.(\ref{EQ5},\ref{EQ14}), we obtain the first-order solution for $\varrho^{(1)}_{ab}$ as
\begin{equation}\label{EQ16}
\begin{split}
\varrho^{(1)}_{ab}=-i\Omega_{b}&\left\{\frac{\gamma_{c}\Lambda(\gamma_{b}-\gamma_{a})\Omega^{2}_{a}}{\left(\Gamma_{cb}\Gamma_{ab}+\Omega^{2}_{a}\right)\mathcal{D}}\right.\\
&\left.+ \frac{(\gamma_{c}\Lambda-\gamma_{b}(\gamma_{c}+\tilde{\gamma}_{p}))(\gamma_{a}\Gamma_{ca}+2\Omega^{2}_{a})\Gamma_{cb}}{\left(\Gamma_{cb}\Gamma_{ab}+\Omega^{2}_{a}\right)\mathcal{D}}\right\},
\end{split}
\end{equation}
To observe gain in the probe transition we obtain the necessary condition as
\begin{equation}\label{EQ17}
(\gamma_{c}\Lambda-\gamma_{b}(\gamma_{c}+\tilde{\gamma}_{p}))(\gamma_{a}\Gamma_{ca}+2\Omega^{2}_{a})\Gamma_{cb}+\gamma_{c}\Lambda(\gamma_{b}-\gamma_{a})\Omega^{2}_{a} >0.
\end{equation}
For large $\Omega_{a} (\gg \gamma_{a}, \gamma_{b},\gamma_{c})$ Eq.(\ref{EQ17}) reduces to
\begin{equation}\label{EQ18}
\gamma_{c}> \gamma_{b}(1+\gamma_{p}/\Lambda).
\end{equation}
Now to observe gain without population inversion Eqs.(\ref{EQ15},\ref{EQ18}) should be satisfied simultaneously. Thus we can summarize the condition for amplification without population inversion in steady-state as
\begin{equation}\label{EQ19}
\begin{cases} 1+\gamma_{p}/\Lambda< \gamma_{c}/\gamma_{b}< \gamma_{p}/(2\Lambda-\gamma_{b}) & \text{if $\Lambda > \gamma_{b}/2$,}
\\
1+\gamma_{p}/\Lambda< \gamma_{c}/\gamma_{b} &\text{if $\Lambda \le \gamma_{b}/2$.}
\end{cases}
\end{equation}
When $\Lambda \le \gamma_{b}/2$, there is no upper limit for $\gamma_{c}$ for  which the probe transition $|a\rangle \leftrightarrow |b\rangle$ will exhibit amplification without population inversion in steady state. From Fig. (2) we observe that, the probe transition $|a\rangle \leftrightarrow |b\rangle$ can exhibit amplification without population inversion even when we have a symmetric bi-directional incoherent pump.
\begin{figure}[t]
\includegraphics[height=4.8cm,width=0.46\textwidth,angle=0]{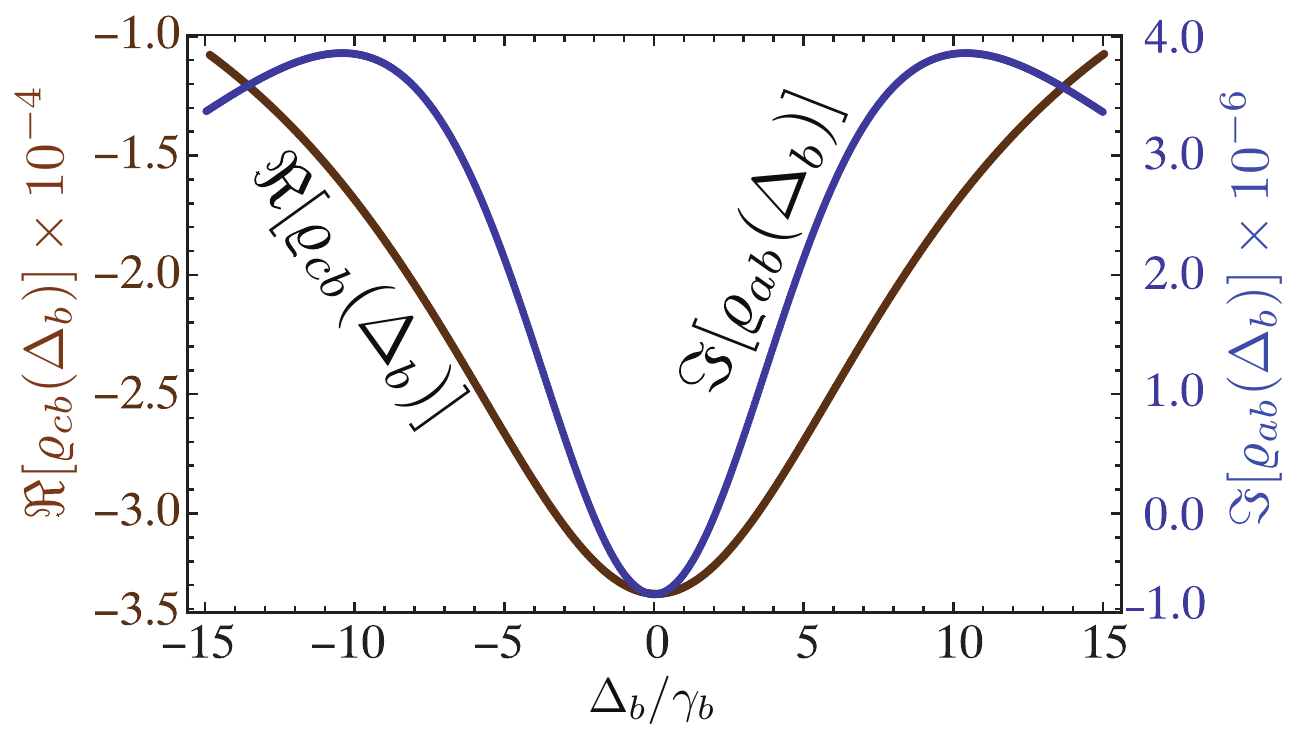}
 \caption{ Effect of the probe detuning on the gain $\propto \Im[\varrho_{ab}(\Delta_{b})]$. For numerical simulation we took $\Omega_{a}=10, \Omega_{b}=0.01, \gamma_{a}=0.01, \gamma_{p}=0.3, \gamma_{b}=1,\gamma_{c}=2,\Lambda=0.5, \Delta_{c}=0$. }
\end{figure}
\begin{figure}[b]
\includegraphics[height=4.6cm,width=0.42\textwidth,angle=0]{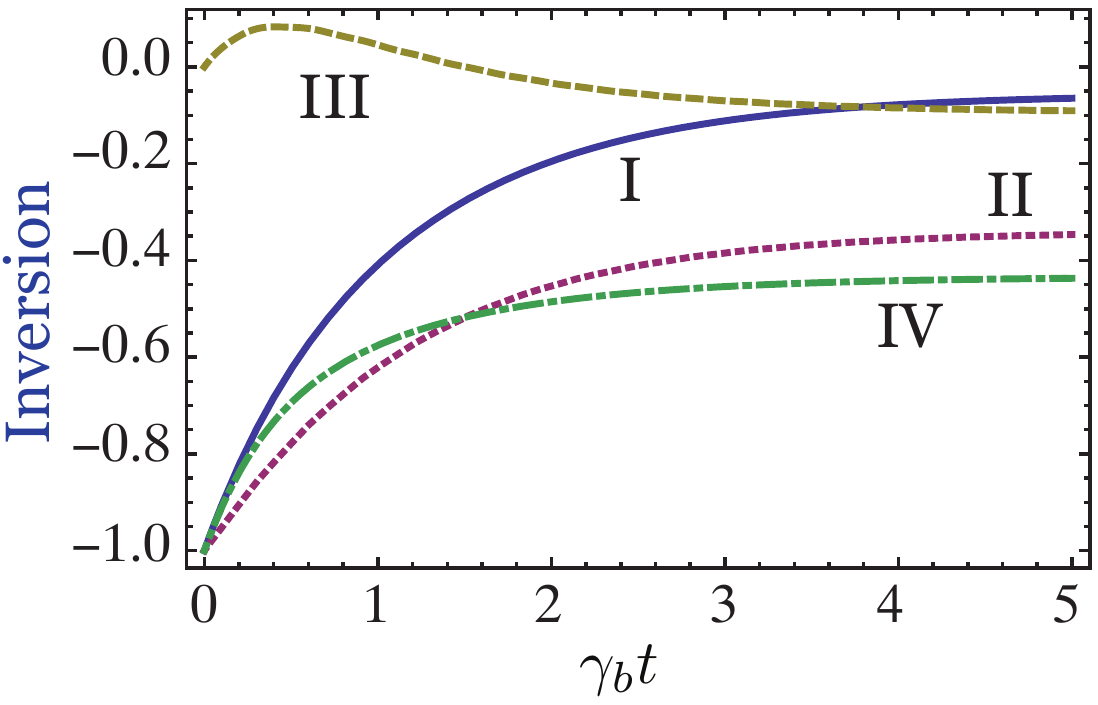}
 \caption{Plots for the transient behavior population inversion in different transitions for bi-directional incoherent pump in the transition $|b\rangle \leftrightarrow |p\rangle$ for the four-level model Fig 1(b). Curve(I): $\varrho_{aa}(t)+\varrho_{cc}(t)+\varrho_{pp}(t)-\varrho_{bb}(t)$, curve (II): $\varrho_{aa}(t)-\varrho_{bb}(t)$, curve (III): $\varrho_{pp}(t)-\varrho_{cc}(t)$ and curve (IV): $\varrho_{pp}(t)-\varrho_{bb}(t)$. For numerical simulation we took $\Omega_{a}=10, \Omega_{b}=0.01, \gamma_{a}=0.01, \gamma_{p}=0.3, \gamma_{b}=1,\gamma_{c}=2,\Lambda=0.5$.}
\end{figure}
\begin{figure*}[htb]
\includegraphics[height=3.8cm,width=0.99\textwidth,angle=0]{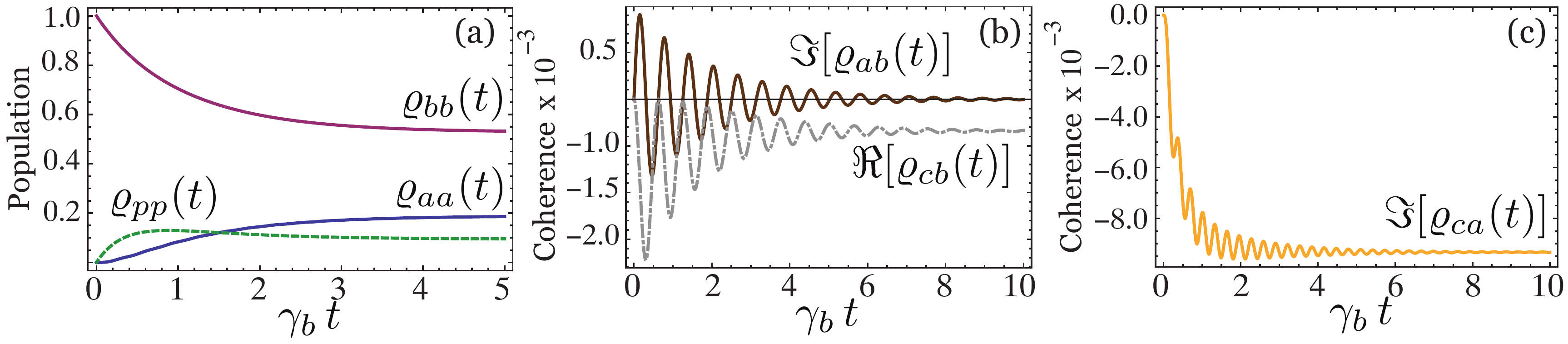}
 \caption{Transient behavior of the population in each levels and the coherences for the bi-directional incoherent pump in the transition $|b\rangle \leftrightarrow |p\rangle$ for the four-level model Fig 1(b). For numerical simulation we took $\Omega_{a}=10, \Omega_{b}=0.01, \gamma_{a}=0.01, \gamma_{p}=0.3, \gamma_{b}=1,\gamma_{c}=2,\Lambda=0.5$.}
\end{figure*}
\noindent Till now we have analyzed the steady-state behavior of the four-level mode when both the probe and driving field are resonant with the corresponding transition. To study the effect of detuning on the gain, let us solve the Eq.(\ref{EQ11}) numerically and the results are shown in Fig.(3). Here we have plotted the steady-state value of $\Im[\varrho_{ab}]$ and $\Re[\varrho_{cb}]$ as a function of probe detuning and resonant drive $(\Delta_{a}=0)$. For the parameters used in Fig. 3, the probe transition will exhibit gain $(\propto \Im[\varrho_{ab}])$ till $\Delta_{b} \sim 3\gamma_{b}$.

To study the evolution of the population and the coherences we will now consider the transient behavior of the four-level medium. Another reason to study the transient regime is that the temporal behavior of the coherence $\varrho_{ab}(t)$ gives us the information about the time interval in which the probe transition will exhibit gain(absorption). This information is readily used in transient lasing without inversion. The probe or the seed pulse is properly delayed so that it enters and cross the medium when the probe transition is exhibiting amplification.
\subsection{Transient state analysis.}
In this section we will show some light on the transient behavior of the four-level model and the main results are shown in Figs. (4,5).  In Fig. 4 we have plotted the transient behavior of the population inversion in different transitions. In steady state, we see that for the parameters used in the numerical simulations we will never observe population inversion i.e $\varrho_{aa}+\varrho_{cc}+\varrho_{pp}-\varrho_{bb} <0$. Infact the system is never inverted at any instant of time (see curve I Fig. 4). Inversion is only observed in the transition $|p\rangle \leftrightarrow |c\rangle$ till $\gamma_{b}t \sim 2$. In Fig. 5(a) we have plotted the temporal evolution of the populations of all the levels. Although the population of the ground state decreased monotonically, the population in level $|p\rangle$ monotonically increases and reaches the steady state after time $\gamma_{b}t \sim 1$. The populations in the level $|a\rangle$ shows some oscillatory behavior, but for the particular choice of the parameters used for numerical simulations, the amplitude of these oscillations are very small. The population in the level $|c\rangle$ closely follow $\varrho_{aa}(t)$. In Fig. 5(b) we have plotted the imaginary part of $\varrho_{ab}(t)$ denoted by $\Im[\varrho_{ab}(t)]$ and the real part of $\varrho_{cb}(t)$ denoted by $\Re[\varrho_{cb}(t)]$. The behavior of $\Im[\varrho_{ab}(t)]$ is also oscillatory. When $\Im[\varrho_{ab}(t)] >0$, the probe transition goes through absorption while for $\Im[\varrho_{ab}(t)]<0$ it exhibits amplification. At resonant drive and probe excitation, the coherence $\varrho_{ab}$ is purely imaginary consequently from Eq. \ref{EQ13}(b) we obtain,
\begin{equation}\label{EQ20}
\Im[\varrho_{ab}(t)]=\frac{1}{2\Omega_{b}}\left[\gamma_{b}\varrho_{aa}-\Lambda \rho_{bb}+\tilde{\gamma}_{p}\varrho_{pp}-\frac{\partial \varrho_{bb}}{\partial t}\right].
\end{equation}
Now the condition for amplification of the probe field $\Im[\varrho_{ab}(t)] <0$ in the transient regime can also be written as
\begin{equation}\label{EQ21}
\frac{\partial \varrho_{bb}}{\partial t} > \gamma_{b}\varrho_{aa}-\Lambda \varrho_{bb}+\tilde{\gamma}_{p}\varrho_{pp}.
\end{equation}
Thus the transient gain condition Eq.(\ref{EQ20}) involves only the population dynamics of the levels $|a\rangle, |b\rangle$ and $|p\rangle$. In short we can say the probe field observe transient gain when the growth of the ground-state population exceeds combined effect of the atoms entering, per unit time,  (due to incoherent decay $\gamma_{b}\varrho_{aa}+\tilde{\gamma}_{p}\varrho_{pp}$) and leaving (due to incoherent pumping to the level $|p\rangle$ given by the rate $\Lambda\varrho_{bb}$)~\cite{TLWI,TLW2}. 

\section{conclusion}
To conclude, in this paper we studied the possibility of steady state amplification without inversion in a four level medium using a symmetric and bi-directional incoherent pump. The four-level model studied here can be conceived as an equivalent  three-level model in cascade configuration with a effective uni-directional pumping needed for steady-state gain. If we consider a bi-directional symmetric pumping in the transition $|c\rangle \leftrightarrow |b\rangle$ the probe transition does not exhibit gain. Though this is true for $\Xi$-configuration, $\Lambda$-configuration $can$ exhibit gain under such circumstances.  In the steady state regime, we found the range of the parameters needed to achieve the amplification without inversion as shown in Fig. 2(a,b,c). We have also briefly highlighted the transient behavior of the system and observed that for the parameters considered here for amplification without inversion in steady-state, the system never shows population inversion at any instant of time. 

\section{Acknowledgement}
We thank M.O.Scully, G.R.Welch, Y.V.Rostovtsev and S.Suckewer for useful discussions and gratefully acknowledge the support from Herman F. Heep and Minnie Belle Heep Texas A$\&$M University Endowed Fund held/administered by the Texas A$\&$M Foundation.
\appendix
\section{Three-Level $\Lambda$-configuration: Gain with bi-directional pump}
The atom-field Hamiltonian in the interaction picture can be written as $(\hbar=1)$
\begin{equation}\label{A1}
\mathcal{V}=-\Delta_{c}|c\rangle\langle c|-\Delta_{b}|b\rangle\langle b|-\left(\Omega_{b}\left |a \rangle \langle b \right |+\Omega_{c}\left |a \rangle \langle c \right | +\text{H.c}\right).
\end{equation}
\noindent Here $\Omega_{c}$ and $\Omega_{b}$ are the Rabi frequencies of the driving and the probe field respectively. We define the detuning $\Delta_{c} =\omega_{ac}-\nu_{c}$ and $\Delta_{b}=\omega_{ab}-\nu_{b}$. In this model the decay in the channels $ab$, $ac$ are quantified by the parameter $\gamma_{b}$,  $\gamma_{c}$ respectively. Incorporating these decay rates, the equation of motion for the atomic density matrix is given as  $(\hbar=1)$
\begin{equation}\label{A2}
\begin{split}
&\frac{\partial \varrho}{\partial t}=-i[\mathcal{V},\varrho]+\frac{\gamma_{b}}{2}\left([\sigma_{b},\varrho\sigma_{b}^{\dagger}]+[\sigma_{b}\varrho,\sigma_{b}^{\dagger}]\right)\\
&+\frac{\gamma_{c}}{2}\left([\sigma_{c},\varrho\sigma_{c}^{\dagger}]+[\sigma_{c}\varrho,\sigma_{c}^{\dagger}]\right)+\frac{\lambda}{2}\left([\sigma_{\lambda},\varrho\sigma_{\lambda}^{\dagger}]+[\sigma_{\lambda}\varrho,\sigma_{\lambda}^{\dagger}]\right)\\
&+\frac{\Lambda}{2}\left([\sigma^{\dagger}_{\lambda},\varrho\sigma_{\lambda}]+[\sigma_{\lambda}^{\dagger}\varrho,\sigma_{\lambda}]\right),
 \end{split}
\end{equation}
where,
\begin{equation}\label{A3}
\begin{split}
 \sigma_{a}=\left |c \rangle \langle a \right |,\, \sigma_{b}=\left |b \rangle \langle a \right |,\,  \sigma_{\lambda}=\left |b \rangle \langle c \right |, \\
 \sigma_{a}^{\dagger}=\left |a \rangle \langle c \right |,\, \sigma_{b}^{\dagger}=\left |a \rangle \langle b \right |,\, \sigma_{\lambda}^{\dagger}=\left |c \rangle \langle b \right |.
 \end{split}
\end{equation}
\begin{figure}[t]
\centerline{\includegraphics[height=4.6cm,width=0.36\textwidth,angle=0]{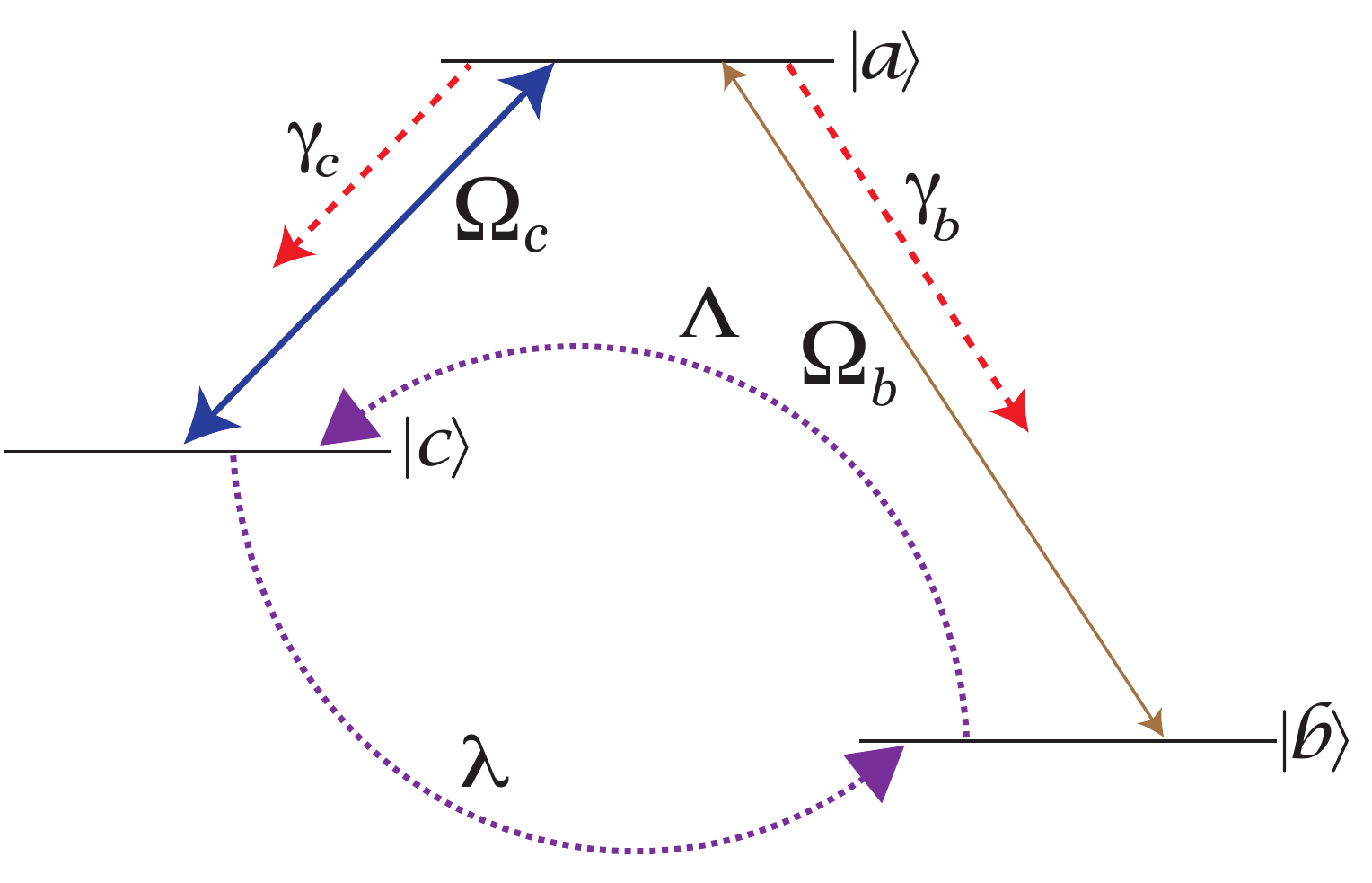}}
 \caption{Three-level model in $\Lambda$-configuration . The spontaneous decay rates $|a\rangle \rightarrow |b\rangle$ and $|a\rangle \rightarrow |c\rangle$ are give by $\gamma_{b}, \gamma_{c}$ respectively. We have introduced an incoherent bi-directional pump in the transition $|b\rangle \leftrightarrow |c\rangle$. The pumping rates are $\lambda (|c\rangle \rightarrow |b\rangle)$ and $\Lambda(|b\rangle \rightarrow |c\rangle)$. $\Omega_{c}$ and $\Omega_{b}$ are the drive and probe field Rabi frequencies.}
\end{figure}
\noindent The equation of motion of the density matrix elements $\varrho_{ll}$ and $\varrho_{ca}$ are given by (for real $\Omega_{c}, \Omega_{b}$)
\begin{subequations}\label{A4}
\begin{align}
\frac{\partial \varrho_{aa}}{\partial t}&=-(\gamma_{b}+\gamma_{c})\varrho_{aa}-2\Omega_{c}\Im[\varrho_{ca}]+2\Omega_{b}\Im[\varrho_{ab}],\\
\frac{\partial \varrho_{bb}}{\partial t}&=\gamma_{b}\varrho_{aa}-\Lambda \rho_{bb}+\lambda\varrho_{cc}-2\Omega_{b}\Im[\varrho_{ab}],\\
\frac{\partial \varrho_{cc}}{\partial t}&=\gamma_{c}\varrho_{aa}-\lambda\varrho_{cc}+\Lambda\varrho_{bb}-\lambda\varrho_{bb}+2\Omega_{c}\Im[\varrho_{ca}],\\
\frac{\partial \varrho_{ca}}{\partial t}&=-\Gamma_{ca}\varrho_{ca}-i\Omega_{c}(\varrho_{cc}-\varrho_{aa})-i\Omega^{*}_{b}\varrho_{cb},
\end{align}
\end{subequations}
where $\Gamma_{ab}=(\gamma_{c}+\gamma_{b}+\Lambda)/2, \Gamma_{ac}=(\gamma_{c}+\gamma_{b}+\lambda)/2, \Gamma_{cb}=(\lambda+\Lambda)/2$. Similar to the earlier discussion, at resonance and in the zeroth order approximation for the probe field we obtained for the steady-state populations
\begin{subequations}\label{A5}
\begin{align}
\varrho^{(0)}_{aa}&=\frac{2\Omega^{2}_{c}\Lambda}{\mathcal{Q}},\\
\varrho^{(0)}_{bb}&=\frac{(\gamma_{b}+\gamma_{c})\lambda\Gamma_{ca}+2(\gamma_{b}+\lambda)\Omega^{2}_{c}}{\mathcal{Q}},\\
\varrho^{(0)}_{cc}&=\frac{(\gamma_{b}+\gamma_{c})\Lambda\Gamma_{ca}+2\Lambda\Omega^{2}_{c}}{\mathcal{Q}},
\end{align}
\end{subequations}
where $\mathcal{Q}=(\gamma_{b}+\gamma_{c})(\lambda+\Lambda)\Gamma_{ca}+2(\gamma_{b}+\lambda+2\Lambda)\Omega^{2}_{c}$. Expression for $\varrho_{ab}$ is the same as Eq.(\ref{EQ5}) with $\Omega_{a} \rightarrow \Omega_{c}$. Using Eqs.(\ref{A5}), we obtain the first-order solution for $\varrho^{(1)}_{ab}$
\begin{equation}\label{A6}
\begin{split}
\varrho^{(1)}_{ab}&=-i\Omega_{b}\left\{\frac{-(\gamma_{b}+\gamma_{c})\lambda\Gamma_{ca}}{\left(\Gamma_{cb}\Gamma_{ab}+\Omega^{2}_{c}\right)\mathcal{Q}}\, +\right.\\
&\left. \frac{\left[(\gamma_{b}+\gamma_{c}+2\Gamma_{cb})\Lambda-2(\gamma_{b}+\lambda)\Gamma_{cb}\right]\Omega^{2}_{c}}{\left(\Gamma_{cb}\Gamma_{ab}+\Omega^{2}_{c}\right)\mathcal{Q}}\right\}.
\end{split}
\end{equation}
To observe gain in the probe transition requires (for large $\Omega_{c}$)
\begin{equation}\label{A7}
\Lambda^{2}-\lambda^{2}+\Lambda\gamma_{c}-\lambda\gamma_{b} >0.
\end{equation}
For symmetric bi-directional pump $(\Lambda =\lambda)$, Eq. (\ref{A7}) reduces to $\gamma_{c} > \gamma_{b}$ which is also the necessary condition for lasing without inversion. Thus for $\gamma_{c} >\gamma_{b}$, the $\Lambda$-configuration exhibits gain in the probe transition in the presence of a symmetric bi-directional pump between the lower two levels. No population inversion $\left(\varrho^{0}_{aa} +\varrho^{0}_{cc}-\varrho^{0}_{bb} <0\right)$ requires
\begin{equation}\label{A8}
(\gamma_{b}+\gamma_{c})(\lambda-\Lambda)\Gamma_{ca}+2(\gamma_{b}+\lambda-2\Lambda)\Omega^{2}_{c} >0.
\end{equation}
For symmetric bi-directional pump, Eq. (\ref{A8}) reduces to $\gamma_{b} > \lambda$ which is the same for three-level model in $\Xi$-configuration. We see that for $\Lambda$-configuration the probe transition can exhibit amplification in the presence of the symmetric bi-directional pump unlike the $\Xi$-configuration.

\section{Three-Level $\Xi$-configuration: Gain with uni-directional pump}
\begin{figure}[h]
\centerline{\includegraphics[height=4.0cm,width=0.20\textwidth,angle=0]{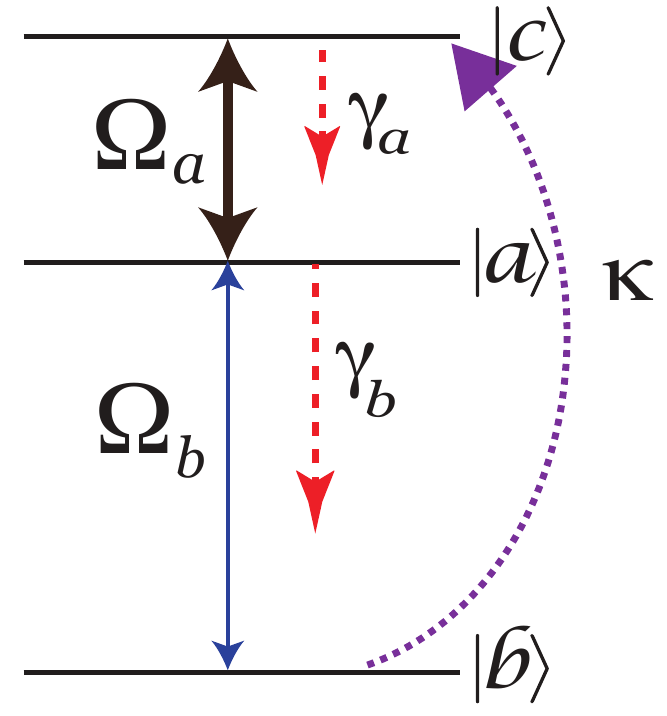}}
 \caption{Three-level model in $\Xi$-configuration . The spontaneous decay rates $|a\rangle \rightarrow |b\rangle$ and $|c\rangle \rightarrow |a\rangle$ are given by $\gamma_{b}, \gamma_{a}$ respectively. We have introduced an incoherent uni-directional pump in the transition $|b\rangle \rightarrow |c\rangle$. The pumping rate is $\kappa$. $\Omega_{a}$ and $\Omega_{b}$ are the drive and probe field Rabi frequencies.}
\end{figure}
To the zeroth order approximation in the probe field we obtained for the steady-state populations (resonant drive and probe excitation)
\begin{subequations}\label{C1}
\begin{align}
\varrho^{(0)}_{aa}&=\frac{\kappa(\gamma_{a}\Gamma_{ca}+2\Omega^{2}_{a})}{\mathcal{R}},\\
\varrho^{(0)}_{bb}&=\frac{\gamma_{b}(\gamma_{a}\Gamma_{ca}+2\Omega^{2}_{a})}{\mathcal{R}},\\
\varrho^{(0)}_{cc}&=\frac{\kappa(\gamma_{b}\Gamma_{ca}+2\Omega^{2}_{a})}{\mathcal{R}},
\end{align}
\end{subequations}
where $\mathcal{R}=\kappa(\gamma_{b}\Gamma_{ca}+2\Omega^{2}_{a})+(\gamma_{b}+\kappa)(\gamma_{a}\Gamma_{ca}+2\Omega^{2}_{a}).$ In the strong field limit $\Omega_{a} \gg \gamma_{b},\gamma_{a},\kappa$, we obtain the non-inversion condition as 
\begin{equation}\label{C2}
\kappa < \gamma_{b}
\end{equation}
Also for gain we obtain
\begin{equation}\label{C3}
\kappa > \sqrt{\gamma_{b}\gamma_{a}}
\end{equation}
Combining Eq.(\ref{C2},\ref{C3}), the condition for gain without population inversion gives~\cite{Shuker1}
\begin{equation}
\sqrt{\gamma_{a}\gamma_{b}} < \kappa <\gamma_{b},\,\, \text{and}\,\,\, \gamma_{b} > \gamma_{a}
\end{equation}
\section{Calculation of population and coherence for the four-level model}
\noindent From Eq. 4(c) we obtain,
\begin{equation}\label{B1}
\varrho^{(0)}_{ca}=-i\frac{\Omega_{a}}{\Gamma_{ca}}\left[\varrho^{(0)}_{cc}-\varrho^{(0)}_{aa}\right].
\end{equation}
From Eq. 13(a) (for real $\Omega_{a}$), we obtain,
\begin{equation}\label{B2}
\varrho^{(0)}_{aa}=\left(\frac{\gamma_{a}}{\gamma_{b}}\right)\varrho^{(0)}_{cc}-2\left(\frac{\Omega_{a}}{\gamma_{b}}\right)\Im[\varrho^{(0)}_{ca}].
\end{equation}
Combining Eq.(\ref{B1},\ref{B2}) we get,
\begin{equation}\label{B3}
\varrho^{(0)}_{cc}=\left[\frac{\gamma_{b}\Gamma_{ca}+2\Omega^{2}_{a}}{\gamma_{a}\Gamma_{ca}+2\Omega^{2}_{a}}\right]\varrho^{(0)}_{aa}.
\end{equation}
From Eq. 13(c), we obtain,
\begin{equation}\label{B4}
\varrho^{(0)}_{pp}=\left(\frac{\gamma_{a}}{\gamma_{c}}\right)\varrho^{(0)}_{cc}-2\left(\frac{\Omega_{a}}{\gamma_{c}}\right)\Im[\varrho^{(0)}_{ca}].
\end{equation}
Combining Eq.(\ref{B1}-\ref{B4}) we get,
\begin{equation}\label{B5}
\varrho^{(0)}_{pp}=\left[\frac{\gamma_{b}}{\gamma_{c}}\right]\varrho^{(0)}_{aa}.
\end{equation}
From Eq. 13(b), we obtain,
\begin{equation}\label{B6}
\varrho^{(0)}_{bb}=\left(\frac{\gamma_{c}+\tilde{\gamma}_{p}}{\Lambda}\right)\varrho^{(0)}_{pp}.
\end{equation}
Combining Eq.(\ref{B5},\ref{B6}) we get,
\begin{equation}\label{B7}
\varrho^{(0)}_{bb}=\left(\frac{\gamma_{c}+\tilde{\gamma}_{p}}{\Lambda}\right)\left[\frac{\gamma_{b}}{\gamma_{c}}\right]\varrho^{(0)}_{aa}.
\end{equation}
Using conservation of the population we obtain the zeroth-order population
\begin{subequations}\label{B8}
\begin{align}
\varrho^{(0)}_{aa}&=\frac{\gamma_{c}\Lambda(\gamma_{a}\Gamma_{ca}+2\Omega^{2}_{a})}{\mathcal{D}},\\
\varrho^{(0)}_{bb}&=\frac{\gamma_{b}(\gamma_{c}+\tilde{\gamma}_{p})(\gamma_{a}\Gamma_{ca}+2\Omega^{2}_{a})}{\mathcal{D}},\\
\varrho^{(0)}_{cc}&=\frac{\gamma_{c}\Lambda(\gamma_{b}\Gamma_{ca}+2\Omega^{2}_{a})}{\mathcal{D}},\\
\varrho^{(0)}_{pp}&=\frac{\gamma_{b}\Lambda(\gamma_{a}\Gamma_{ca}+2\Omega^{2}_{a})}{\mathcal{D}},
\end{align}
\end{subequations}
where, $\mathcal{D}=2\Omega^{2}_{a}\left[2\Lambda\gamma_{c}+\gamma_{b}(\Lambda+\gamma_{c}+\tilde{\gamma}_{p})\right] +\gamma_{a}\Gamma_{ca}\left[\gamma_{c}\Lambda+\gamma_{b}(\Lambda+\gamma_{c}+\tilde{\gamma}_{p})\right]+\gamma_{c}\gamma_{b}\Lambda\Gamma_{ca}$. The coherence $\varrho^{(0)}_{ca}$ is given as
\begin{equation}\label{B9}
\varrho^{(0)}_{ca}=-i\left[\frac{\Omega_{a}\gamma_{c}\Lambda(\gamma_{b}-\gamma_{a})}{\mathcal{D}}\right].
\end{equation}
Using Eq. 4(c)  and Eqs. (\ref{B8}), we can also easily obtain an analytical expression for $\varrho^{(1)}_{cb}$. Using similar line of action we can also obtain the populations and coherences for the three-level models in $\Lambda$ and $\Xi$ configurations and the results are used in the text.


\begin{thebibliography}{99}
\bibitem{D} Some authors also use the definition of detuning as $\Delta =\nu-\omega$ where $\nu$ and $\omega$ are the drive and the atomic transition frequencies respectively.
\end{thebibliography}
\end{document}